\documentclass[aps,prd,twocolumn]{revtex4-2}
\usepackage[colorlinks=true, linkcolor=blue, citecolor=blue, urlcolor=blue]{hyperref}
\usepackage{bm,latexsym,amssymb,amsmath,mathrsfs}
\usepackage{graphicx}
\usepackage{color}
\usepackage{times,setspace}

\begin{document}

\title{Quantum sampling for the Euclidean path integral of lattice gauge theory}
\author{Arata Yamamoto}
\affiliation{Department of Physics, The University of Tokyo, Tokyo 113-0033, Japan}

\begin{abstract}
Although the Hamiltonian formalism is so far favored for quantum computation of lattice gauge theory, the path integral formalism would never be useless.
The advantages of the path integral formalism are the knowledge and experience accumulated by classical lattice simulation and manifest Lorentz invariance.
We discuss quantum computation of lattice gauge theory in the path integral formalism.
We utilize a quantum sampling algorithm to generate gauge configurations, and demonstrate a benchmark test of $Z_2$ lattice gauge theory on a four-dimensional hypercube.
\end{abstract}

\maketitle

\section{Introduction}

Computational devices for lattice gauge theory are on the verge of shifting from classical computers to quantum computers \cite{Zohar:2015hwa,Dalmonte:2016alw,Banuls:2019bmf}.
Quantum computation is expected as a technological breakthrough for unsolved hard problems, e.g., quantum chromodynamics (QCD) at nonzero baryon density \cite{Clemente:2020lpr,Yamamoto:2021vxp}.
For quantum computation of lattice gauge theory, the Hamiltonian formalism is favored because the algorithmic implementation is straightforward.
This is, in some sense, sad news for lattice QCD researchers.
The path integral formalism has been used in lattice QCD for a long period.
A vast number of the data and algorithms they developed are not available for quantum computation.
Another drawback of the Hamiltonian formalism is explicit breaking of the Lorentz invariance.
The difference between temporal and spatial discretization leads to anisotropic renormalization \cite{Carena:2021ltu}.
Simulation parameters must be fine-tuned to trace the line of constant physics.
In contrast, the path integral is isotropic in a four-dimensional spacetime, so manifestly Lorentz invariant.

Since the path integral of lattice gauge theory is multiple integral over classical numbers, it can be regarded as the partition function of a classical statistical system.
There are many proposals for quantum sampling of classical statistical systems \cite{PhysRevE.56.3661,PhysRevLett.99.030603,PhysRevLett.101.130504,PhysRevA.78.042336,PhysRevA.82.060302,PhysRevLett.116.080503,PhysRevLett.127.100504,PhysRevA.104.032602}.
In classical Monte Carlo sampling, configurations are generated by a Markov chain with classical random numbers.
In quantum sampling, configurations are generated by quantum fluctuation.
Quantum sampling algorithms can achieve quadratic speedup over classical Markov-chain algorithms \cite{PhysRevLett.101.130504,PhysRevA.78.042336} or further speedup \cite{PhysRevA.82.060302,PhysRevLett.116.080503,PhysRevLett.127.100504,PhysRevA.104.032602} although the degree of speedup is sensitive to the algorithm and system.
In any case, quantum sampling would be a promising approach to large-scale statistics.

In this paper, we discuss the application of the quantum sampling algorithms to the path integral of lattice gauge theory.
Basics of lattice gauge theory in the path integral formalism are briefly reviewed in Sec.~\ref{sec2}.
We focus on one of the quantum sampling algorithms.
The algorithm is explained in Sec.~\ref{sec3} and the results of a test simulation are shown in Sec.~\ref{sec4}.
The combined use of the quantum sampling algorithm and the quantum adiabatic algorithm successfully reproduces correct results.
While the test was done on the Qiskit noiseless simulator, the method is executable on real quantum devices.
Some comments on practical applications are given in Sec.~\ref{sec5}.

\section{$Z_2$ lattice gauge theory}
\label{sec2}

We mainly consider the $Z_2$ lattice gauge theory without matter fields for the sake of simplicity.
Although the $Z_2$ lattice gauge theory does not have the corresponding continuum theory, it is a good benchmark for quantum algorithms because the mapping of a gauge link variable to a qubit is trivial.
Previous works developed the quantum simulation of the $Z_2$ lattice gauge theory in the Hamiltonian formalism \cite{Zohar:2016wmo,Yamamoto:2020eqi,Gustafson:2020yfe,Gustafson:2021jtq}.
We here overview the path integral formulation of the $Z_2$ lattice gauge theory.

\begin{figure}[h]
\begin{center}
 \includegraphics[width=.48\textwidth]{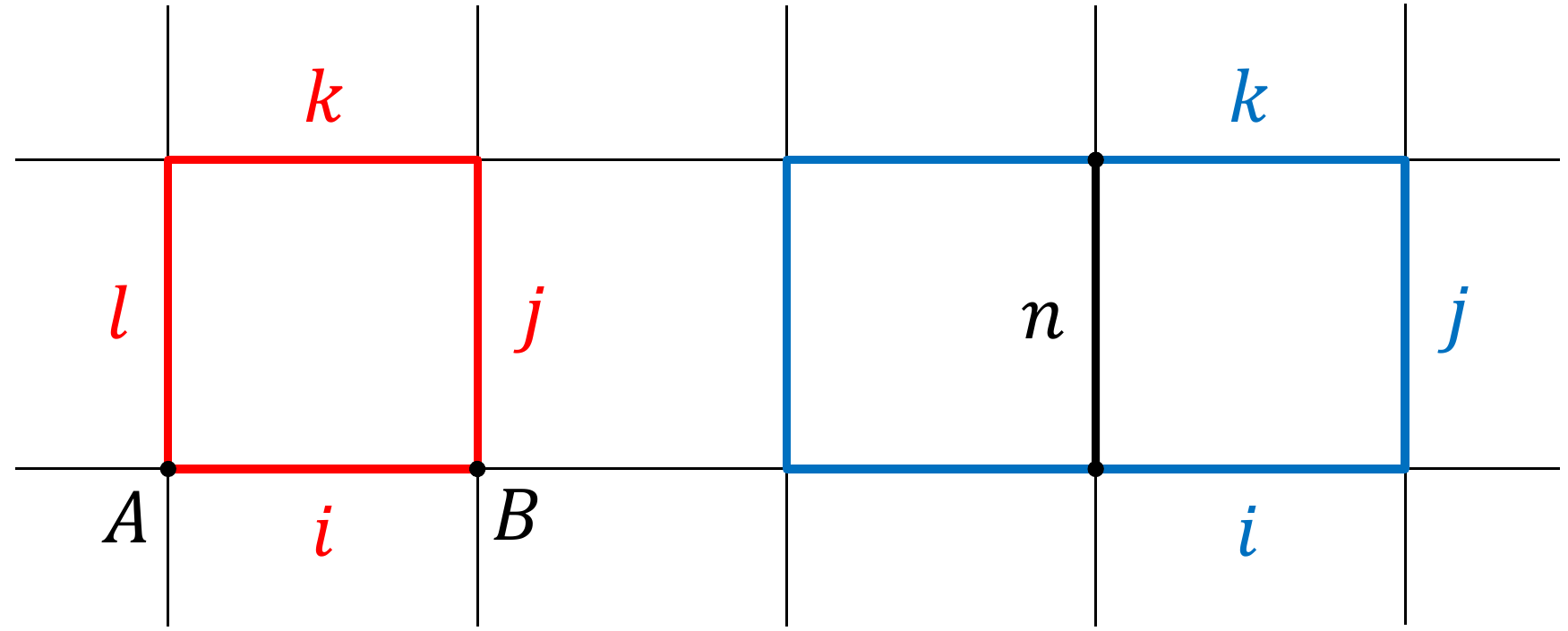}
\caption{
\label{figps}
Plaquette (left) and staples (right).
}
\end{center}
\end{figure}

Let us consider a four-dimensional hypercubic lattice in the Euclidean spacetime $(x,y,z,\tau)$.
Gauge fields are defined as link variables $U_n$ $(n=1,2,\cdots,N)$ on the lattice.
As gauge group is discrete, the path integral is the multiple sum over the link variables 
\begin{equation}
 \mathcal{Z}= \sum_{U_1} \sum_{U_2} \cdots \sum_{U_N} e^{-S}.
\label{eqZc}
\end{equation}
Each link variable takes the classical value $U_n=+ 1$ or $-1$.
A physical observable is a function of the link variables.
The expectation value is given by
\begin{equation}
 \langle O \rangle = \frac{1}{\mathcal{Z}} \sum_{U_1} \sum_{U_2} \cdots \sum_{U_N} O e^{-S}.
\label{eqOc}
\end{equation}
The most familiar form of the classical action is 
\begin{equation}
 S = - \beta \sum_{\{ijkl\}} U_i U_j U_k U_l ,
\end{equation}
where the summation runs over all the plaquettes (Fig.~\ref{figps}).
The parameter $\beta$ is related to the conventional gauge coupling constant $g$ via $\beta=1/g^2$.
In the strong coupling limit $\beta \to 0$ ($g \to \infty$), the link variables are random.
The average value of plaquettes is zero.
In the weak coupling limit $\beta \to \infty$ ($g \to 0$), only the minimum action $S=-\beta N_{\rm plaq}$, where $N_{\rm plaq}$ is the number of plaquettes, survives.
All the plaquettes are unity.

In classical Monte Carlo sampling, $\{U_1,U_2,\cdots,U_N\}$ is stochastically generated with a probability of $e^{-S}$.
This is called the gauge configuration.
Classical computation of lattice gauge theory involves two steps: generating gauge configurations and then calculating the expectation value from the generated gauge configurations.
The latter step does not affect the former step.
Once the gauge configurations are generated, they are stored for a long time, say, for several months or years.
They can be reused for studying other observables.

When the number of links is $N$, the total number of classical states is $2^N$.
This number contains gauge redundancy, so can be reduced by gauge fixing.
The $Z_2$ gauge transformation is described as
\begin{equation}
 U_i \to U'_i = \Lambda_A U_i \Lambda_B.
\end{equation}
$\Lambda_A$ and $\Lambda_B$ are the gauge transformation function at the end points of the $i$-th link (Fig.~\ref{figps}).
They can be arbitrarily chosen as $+1$ or $-1$.
We can fix $U'_i=1$ by taking $U_i=\Lambda_A \Lambda_B$.
In general, when the number of lattice sites is $N_{\rm site}$, $N_{\rm site}-1$ link variables can be fixed by $N_{\rm site}-1$ relative signs of the gauge transformation function.
The number of independent link variables is reduced from $N$ to $N-N_{\rm site}+1$.
Although gauge fixing is not mandatory, it can help reduce memory size.

\section{Algorithms}
\label{sec3}

Among many quantum sampling algorithms, we adopt the algorithm developed in Refs.~\cite{PhysRevLett.127.100504,PhysRevA.104.032602}.
We define the parent Hamiltonian
\begin{equation}
 H = N \left( I-e^{-\frac{\mathcal{S}}{2}} M e^{\frac{\mathcal{S}}{2}} \right).
\label{eqH1}
\end{equation}
This is not a physical Hamiltonian of the system but a working Hamiltonian for quantum simulation.
The Hilbert space is spanned by the $2^N$-dimensional vector $|U_1 \rangle|U_2 \rangle\cdots|U_N \rangle$.
The matrix $\mathcal{S}$ is constructed by encoding the classical action $S$ as the diagonal matrix
\begin{equation}
 \mathcal{S} = - \beta \sum_{\{ijkl\}} Z_i Z_j Z_k Z_l ,
\label{eqqs}
\end{equation}
i.e., $\mathcal{S} |U_1 \rangle|U_2 \rangle\cdots|U_N \rangle = S |U_1 \rangle|U_2 \rangle\cdots|U_N \rangle$.
The matrix $M$ is defined by the matrix representation of a Markov chain.
It is formally written as
\begin{equation}
 M = I - p + p \Gamma.
\end{equation}
This formal equation means that the update $\Gamma$ is accepted with the probability $p$ and rejected with the probability $I-p$.
The choice for the Markov chain is not unique, but any ergodic Markov chain eventually reaches unique equilibrium.
The Perron-Frobenius theorem ensures that the largest eigenvalue of $M$ is unity and the smallest eigenvalue of $H$ is zero.
One can easily find that the ground state is
\begin{equation}
 |\Psi\rangle = \frac{1}{\sqrt{\mathcal{Z}}} \sum_{U_1} \sum_{U_2} \cdots \sum_{U_N} e^{-\frac{\mathcal{S}}{2}} |U_1 \rangle|U_2 \rangle\cdots|U_N \rangle
\label{eqgs}
\end{equation}
and the corresponding eigenvalue is zero.
The expectation value \eqref{eqOc} can be obtained by the matrix representation of $O$ as
\begin{equation}
 \langle O \rangle = \langle\Psi| \mathcal{O} |\Psi\rangle,
\end{equation}
with $\mathcal{O} |U_1 \rangle|U_2 \rangle\cdots|U_N \rangle = O |U_1 \rangle|U_2 \rangle\cdots|U_N \rangle$.

When the Glauber dynamics is chosen as the Markov chain, the Hamiltonian is written in a simple form.
In the Glauber dynamics algorithm, one of the link variables is selected with a probability $1/N$ and its sign is flipped with a probability $e^{-\Delta S}/(1+e^{-\Delta S})$, where $\Delta S$ is the change in the action by the flip.
In the matrix form, the flipping probability for the $n$-th link is 
\begin{equation}
 p_n = \frac{1}{N} \frac{e^{-\beta Z_n C_n}}{2\cosh(\beta C_n)}.
\label{eqpn}
\end{equation}
Here we define the sum of the ``staples''
\begin{equation}
 C_n = \sum_{\{ijk\}} Z_i Z_j Z_k
\end{equation}
connected to the $n$-th link (see Fig.~\ref{figps}).
One link is connected to six staples (except at boundaries) on the four-dimensional lattice.
The Markov-chain matrix is
\begin{equation}
 M = I- \sum_n p_n + \sum_n p_n X_n.
\end{equation}
After some algebra, we get
\begin{equation}
 H = \sum_n \frac12 \left( I- \tanh(\beta C_n) Z_n - \frac{1}{\cosh(\beta C_n)} X_n \right).
\label{eqH2}
\end{equation}
The Hamiltonian is described by the Pauli gates.
This is the simplest choice as for the $Z_2$ gauge group.
When gauge group is continuous, the update should be continuous, so other choices will be better. 
For example, $\Gamma$ is a uniform matrix for the Metropolis algorithm and the molecular-dynamics update for the Hybrid Mote Carlo algorithm.
The probability $p$ is still given by the change in the action by these updates but not local as Eq.~\eqref{eqpn}.

The ground state of the Hamiltonian can be computed by the quantum adiabatic algorithm \cite{Farhi2000,Farhi2001-mi}.
The algorithm is written as the evolution equation
\begin{equation}
 |\Psi(\beta)\rangle = e^{-i\int_0^T dt H (\beta')} |\Psi(\beta_0)\rangle,
\label{eqpsia}
\end{equation}
where $H (\beta')$ is gradually changed from $H (\beta_0)$ at $t=0$ to $H (\beta)$ at $t=T$.
Note that the simulation time $t$ has nothing to do with the physical time $\tau$.
If the initial state $|\Psi(\beta_0)\rangle$ is the ground state of $H(\beta_0)$, the final state $|\Psi(\beta)\rangle$ is the ground state of $H(\beta)$.
The performance of the quantum adiabatic algorithm depends on the spectral gap of $H (\beta')$ along the path from $\beta' = \beta_0$ to $\beta' = \beta$.
The convergence is faster as the gap between the smallest and second smallest eigenvalues is larger.
Since the gap depends on the choice of the Markov-chain matrix and the path, we should choose as good ones as possible.

Another way to compute the ground state is the quantum variational algorithm \cite{Peruzzo2014-hk}.
Although the obtained ground state is an approximate one, the circuit depth is lower than that of the quantum adiabatic algorithm.
The quantum variational algorithm is more robust against noise, so more realistic for the simulation on near-term quantum devices.

\section{Benchmark test}
\label{sec4}

\begin{figure}[h]
\begin{center}
 \includegraphics[width=.48\textwidth]{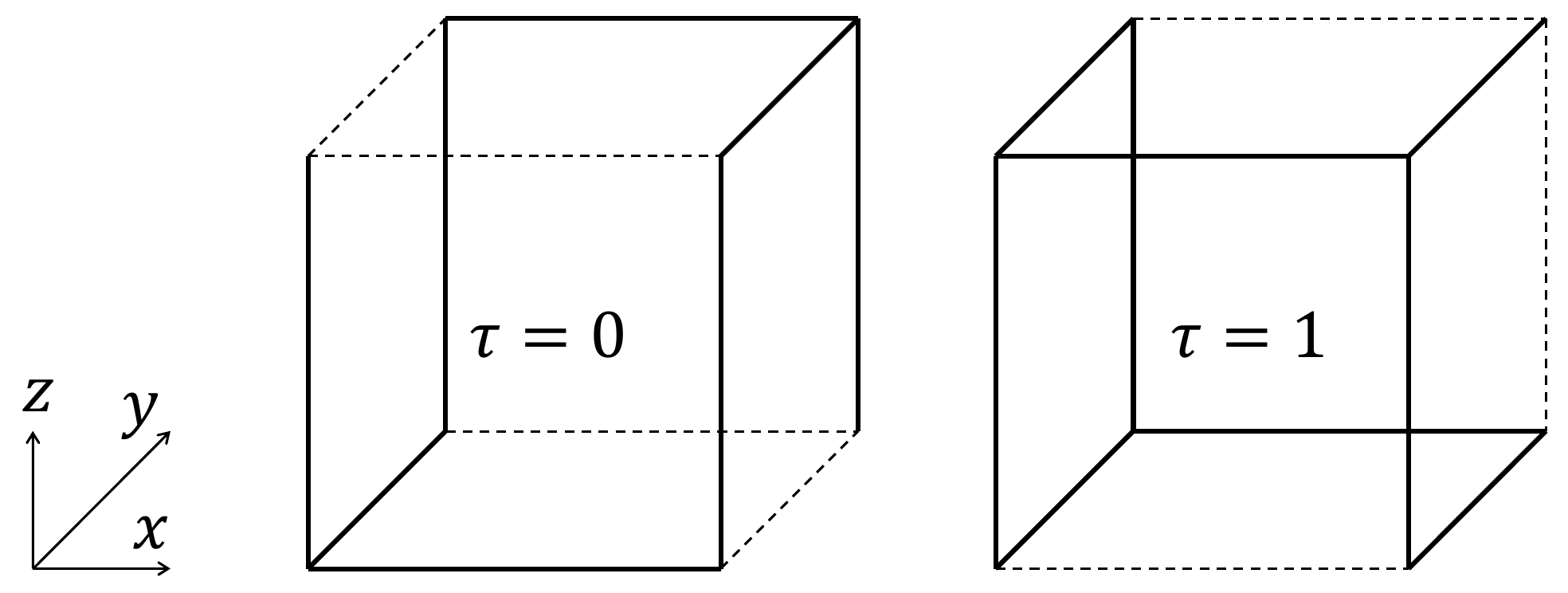}
\caption{
\label{fighc}
Four-dimensional hypercube.
The dotted link variables and all temporal link variables (not shown) are set to $+1$ by gauge fixing.
}
\end{center}
\end{figure}

We ran the benchmark test of one four-dimensional hypercube, i.e., the $2 \times 2 \times 2 \times 2$ lattice with open boundary conditions.
Each link is connected to three staples in this geometry.
The number of link variables is $N=32$, the number of sites is $N_{\rm site}=16$, and the number of plaquettes is $N_{\rm plaq}=24$. 
Fifteen link variables are fixed as shown in Fig.~\ref{fighc}.
Thanks to gauge fixing, the dimension of the Hilbert space is reduced from $2^{32}$ to $2^{17}$.
The minimum action state is $2^{15}$-fold degenerate before gauge fixing but non-degenerate after gauge fixing.

We adopted the quantum adiabatic calculation \eqref{eqpsia} to obtain the ground state of the parent Hamiltonian \eqref{eqH2}.
Two initial states are possible: one is the ``cold start'', the weak coupling limit $\beta_0=\infty$,
\begin{equation}
 |\Psi(\infty)\rangle = |1\rangle |1\rangle \cdots |1\rangle,
\end{equation}
and the other is the ``hot start'', the strong coupling limit $\beta_0 = 0$,
\begin{equation}
 |\Psi(0)\rangle = 2^{-\frac{17}{2}} (|1\rangle+|-1\rangle) (|1\rangle+|-1\rangle) \cdots (|1\rangle+|-1\rangle).
\end{equation}
It is easy to see that the expectation value of the Hamiltonian \eqref{eqH2} is zero for these states.
The evolution operator is approximated by the Suzuki-Trotter decomposition with a small time step $\delta t$.
We introduced ancillary qubits to keep the information of the staple operator $C_n$ and implemented each term of the decomposed evolution operator by the rotation gate $R_Z$ or $R_X$ controlled by the ancillae.
We used $\delta t = 0.2$ and $\delta t =0.5$ and checked the consistency of the results.

\begin{figure}[t]
\begin{center}
 \includegraphics[width=.48\textwidth]{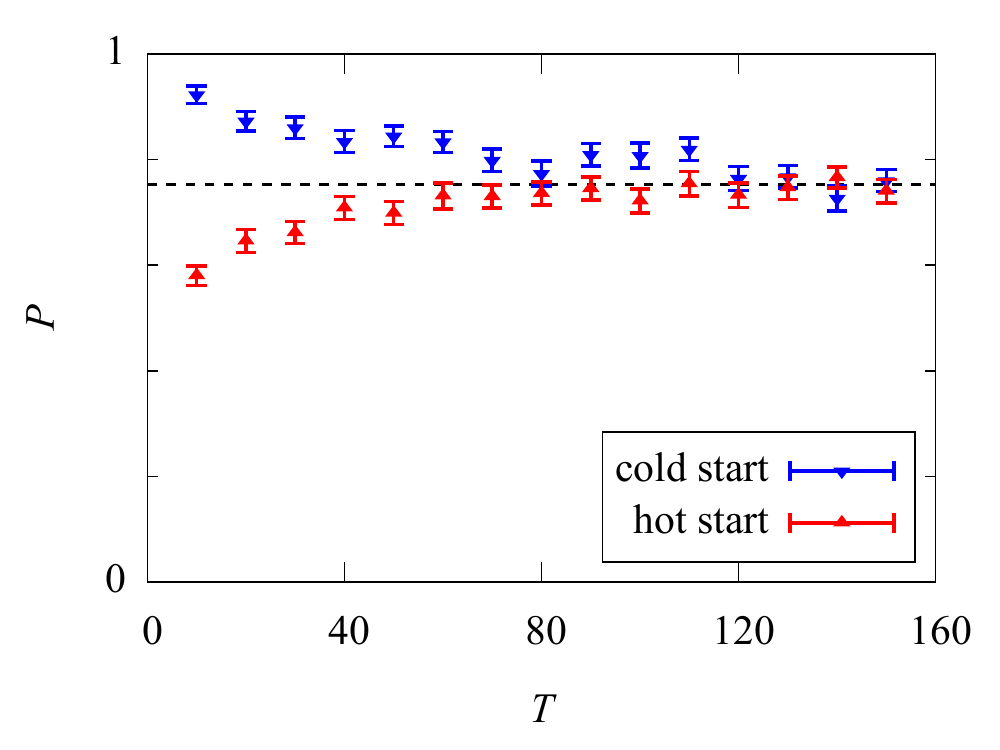}
\caption{
\label{figT}
Plaquette value $P$ as a function of the simulation time $T$ in the quantum adiabatic calculation.
The inverse coupling constant is $\beta=0.7$.
The broken line is the exact value $P\simeq 0.753$.
}
\end{center}
\end{figure}
\begin{figure}[t]
\begin{center}
 \includegraphics[width=.48\textwidth]{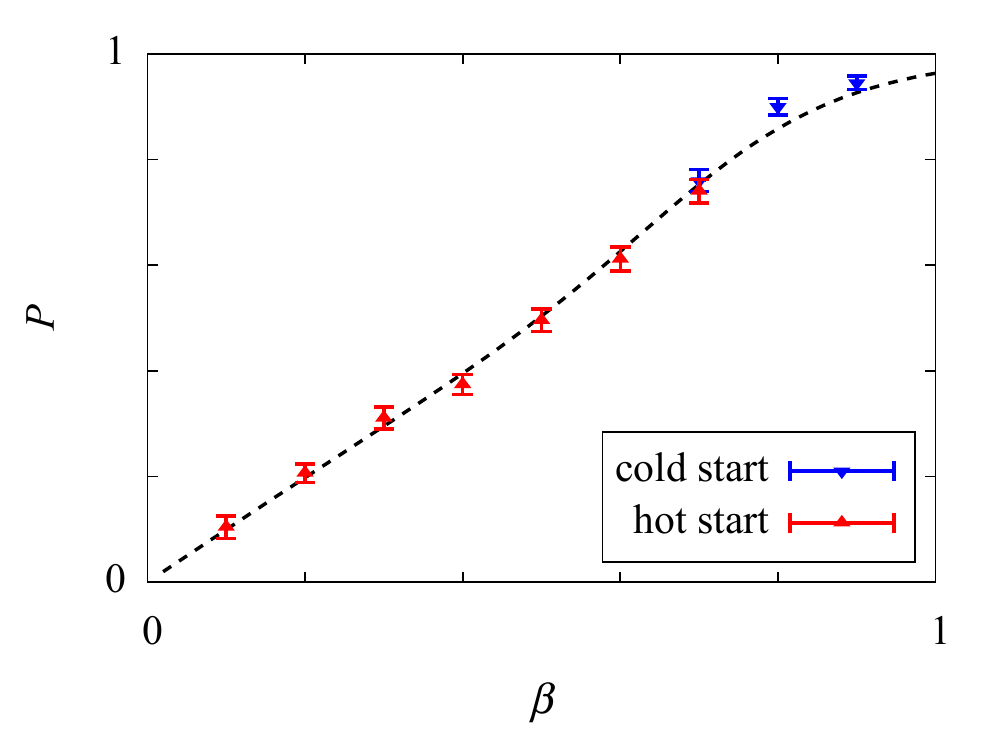}
\caption{
\label{figB}
Plaquette value $P$ as a function of the inverse coupling constant $\beta$.
The broken curve is the exact value.
}
\end{center}
\end{figure}

A typical behavior of the quantum adiabatic calculation is shown in Fig.~\ref{figT}.
The average plaquette value
\begin{equation}
 P = \frac{1}{N_{\rm plaq}} \sum_{\{ijkl\}} \langle \Psi(\beta)| Z_i Z_j Z_k Z_l |\Psi(\beta)\rangle
\end{equation}
is plotted.
The calculations with the cold and hot starts converge to the same value when the simulation time $T$ is large enough.
The exact value of $P$ obtained by the brute-force calculation of Eqs~\eqref{eqZc} and \eqref{eqOc} is also shown.
The converged value is consistent with the exact value.
The converged values for various $\beta$ are summarized in Fig.~\ref{figB}.
The quantum sampling simulation works well in the entire region of $\beta$.
We found that the simulation time to reach the convergence depends on $\beta$ and the initial state.
The hot start is efficient for small $\beta$ and the cold start is efficient for large $\beta$, as expected.

\section{Practical applications}
\label{sec5}

In practice, we repeatedly execute the quantum circuit, measure $\{ U_1, U_2,\cdots, U_N \}$ for each shot, and store them on classical registers.
They are nothing but the gauge configurations.
The expectation value can be calculated from the gauge configurations by conventional programs on classical computers.
Since the gauge configurations are the sets of classical numbers, they can be kept permanently.
This is also an advantage of the path integral formalism.
In the Hamiltonian formalism, quantum states are generated and stored on quantum registers.
Such quantum states can be kept only for a short term and used only for one measurement.
They cannot be reused after the measurement is performed.

There is a difference between classical and quantum sampling procedures.
In the classical Markov-chain Mote Carlo, the first gauge configuration is generated by applying many Markov-chain updates to an initial configuration.
Other gauge configurations are sequentially generated by applying the updates to the previous configurations.
The number of the subsequent updates can be taken to be smaller than the number of the initial updates as far as autocorrelation is harmless.
On the other hand, the quantum circuit must be restarted after each measurement, so does not have such a shortcut.
It is important to find an initial state with good convergence, as discussed in Sec.~\ref{sec4}.

We can show quadratic speedup in the two-dimensional $Z_2$ lattice gauge theory.
When a two-dimensional space-time is large enough, boundaries are negligible and the link variables in one direction are erased by gauge fixing.
The action is reduced to $S=-\beta\sum_{ij} U_i U_j$.
This is equivalent to the classical Ising Hamiltonian, where quadratic speedup is achievable \cite{PhysRevLett.127.100504,PhysRevA.104.032602}.
In the four-dimensional $Z_2$ lattice gauge theory, we need a numerical analysis of how computational cost scales as a function of system size.
From a practical viewpoint, however, pure gauge theory is not so important because the computational cost is mild even on classical computers.
The original motivation for quantum computation is to solve the problem of large computational cost.
The large computational cost comes from the quark part in lattice QCD.
In the conventional formulation, the fermion action is rewritten as a determinant, and then replaced by the integral of a scalar field, which is called the pseudofermion field.
Since the pseudofermion action contains the inverse of the Dirac operator, its calculation is time-consuming.
We must construct a quantum circuit to compute the pseudofermion action or find a more economical formulation of the fermion action.
The scaling property of its computational cost is critical for the practicality of quantum sampling simulation for lattice QCD.

Another intriguing application is the sign problem.
The Dirac determinant is complex when a baryon chemical potential is nonzero.
When the path integral has complex weight, the Markov-chain matrix $M$ is ill-defined.
If $M$ is generalized to a complex matrix, the convergence to a stationary configuration is not validated, so the algorithm in Sec.~\ref{sec3} breaks down.
This can be naively cured by the reweighting method, where a complex phase factor is absorbed into the observable $O$, but the sign problem is still hiding behind.
The reweighted ensemble is very noisy and its statistical error is very large.
The required number of gauge configurations to obtain reasonable accuracy becomes unacceptably large, typically exponentially large.
This is well-known in classical Monte Carlo simulation and will be common to quantum sampling.
The sign problem could be solved only if quantum computation achieves exponential speedup or polynomial speedup with a large power.

\begin{acknowledgments}
The author was supported by JSPS KAKENHI Grant No.~19K03841.
\end{acknowledgments}

\bibliographystyle{apsrev4-2}
\bibliography{paper}

\end{document}